\documentclass{ws-procs975x65}
\usepackage[colorlinks=true,urlcolor=blue,hypertex]{hyperref}

\begin{document}
\def\figsubcap#1{\par\noindent\centering\footnotesize(#1)}

\title{\uppercase{Towards the future of supernova cosmology}}

\author{\uppercase{Michelle Knights}$^{1,2}$, \uppercase{Bruce A. Bassett}$^{1,2,3}$, \uppercase{Melvin Varughese}$^{1,4}$, \uppercase{Ren\'ee Hlozek}$^{5}$, \uppercase{Martin Kunz}$^{6}$, \uppercase{Mat Smith}$^{7}$ and \uppercase{James Newling}$^{1,2}$}

\address{[1] African Institute for Mathematical Sciences, 6 Melrose Road, Muizenberg, 7945, South Africa
\\Email: michelle.knights@gmail.com}
\address{[2] Department of Mathematics and Applied Mathematics, University of Cape Town, Rondebosch, Cape Town, 7700, South Africa}
\address{[3] South African Astronomical Observatory, Observatory Road, Observatory, \\Cape Town, 7935, South Africa}
\address{[4] Department of Statistics, University of Cape Town, Rondebosch, Cape Town, 7700, South Africa}
\address{[5] Department of Astrophysical Sciences, Princeton University, Princeton, NJ 08544, USA}
\address{[6] D\'epartement de Physique Th\'eorique and Center for Astroparticle Physics, Universit\'e de Gen\`eve, Quai E.\ Ansermet 24, CH-1211 Gen\`eve 4, Switzerland} 
\address{[7] Department of Physics, University of the Western Cape, Bellville,\\Cape Town, 7535, South Africa}


\begin{abstract}
 For future surveys \cite{des,pan,lsst}, spectroscopic follow-up for all supernovae will be extremely difficult. However, one can use light curve fitters \cite{salt,mlcs}, to obtain the probability that an object is a Type Ia. One may consider applying a probability cut to the data, but we show that the resulting non-Ia (nIa) contamination can lead to biases in the estimation of cosmological parameters. A different method, which allows the use of the full dataset and results in unbiased cosmological parameter estimation, is Bayesian Estimation Applied to Multiple Species (BEAMS) \cite{kunz}. BEAMS is a Bayesian approach to the problem which includes the uncertainty in the types in the evaluation of the posterior. Here we outline the theory of BEAMS and demonstrate its effectiveness using both simulated datasets and SDSS-II data. We also show that it is possible to use BEAMS if the data are correlated, by introducing a numerical marginalisation over the types of the objects.
\end{abstract}


\section*{An Introduction to BEAMS}
\begin{figure}[th]
 \centering
\parbox{2.4in}{\epsfig{figure=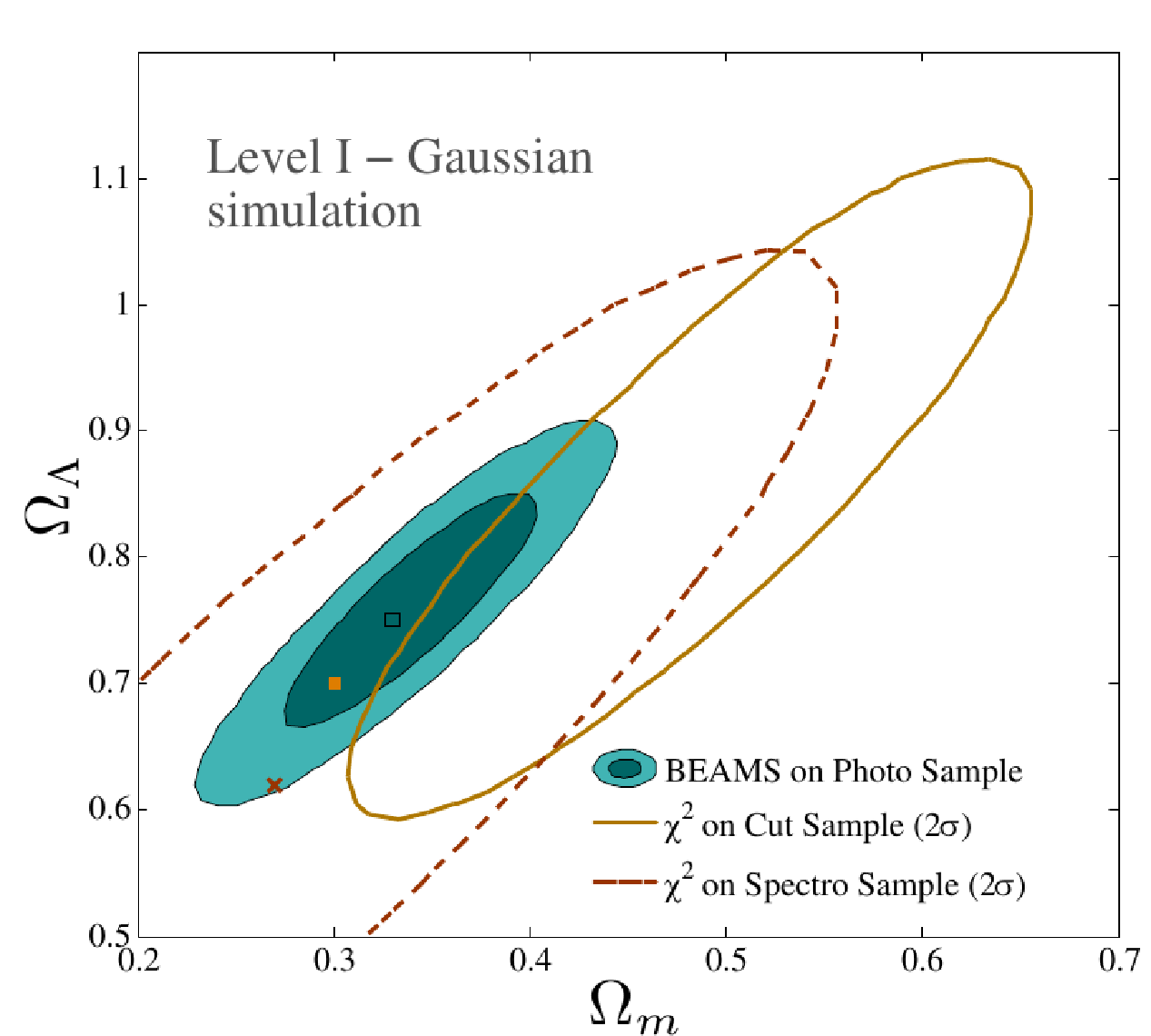,width=2.3in}
\figsubcap{a}}
 \hspace*{4pt}
 \parbox{2.4in}{\epsfig{figure=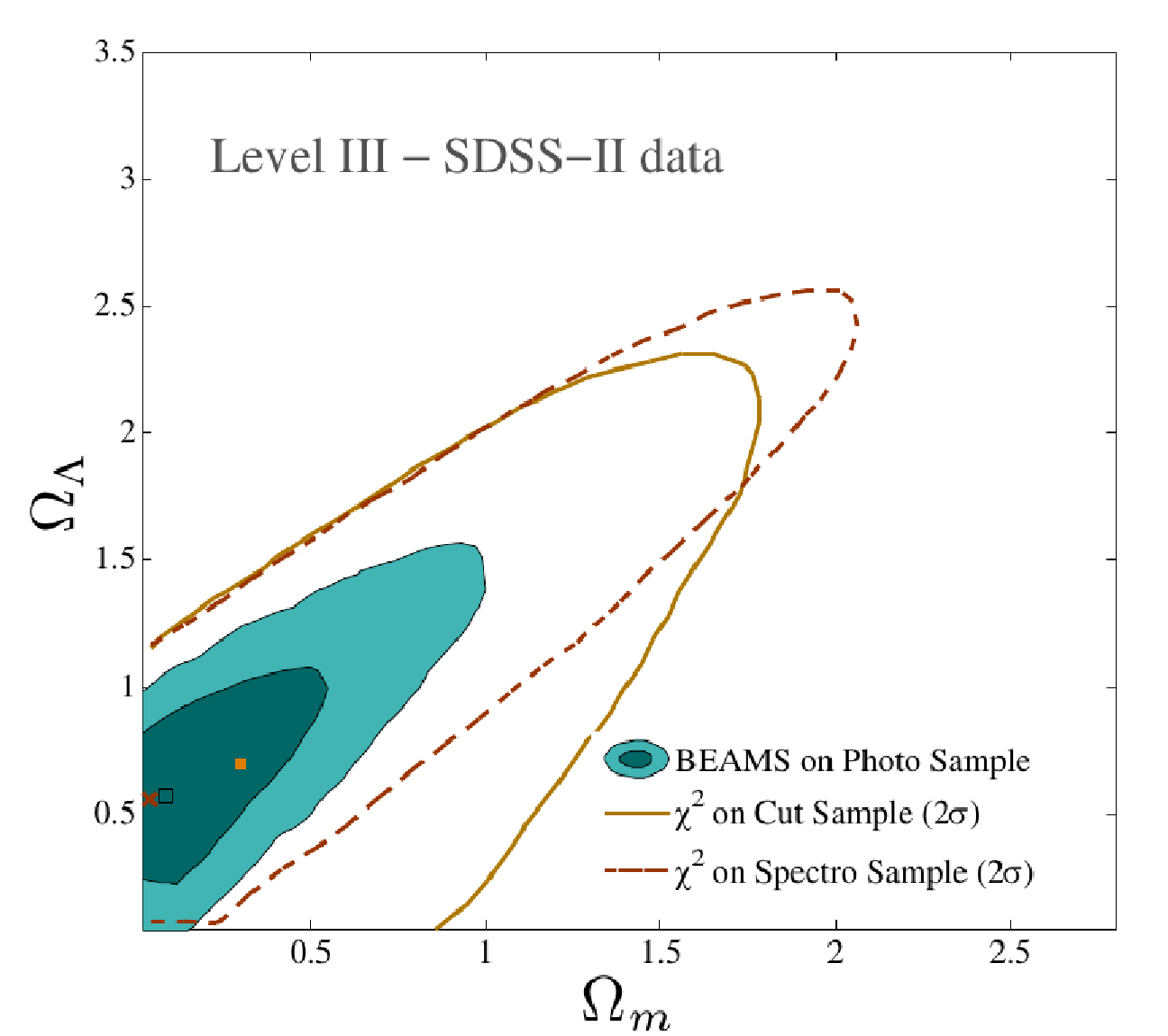,width=2.3in}
 \figsubcap{b}}
\caption{$\Omega_m-\Omega_\Lambda$ contours produced by analysing (a) simulated data and (b) SDSS-II data, using: only the spectroscopically confirmed objects (dashed line), a sample with a probability cut of 0.9 (solid line) and BEAMS (filled contours). The filled square, empty square and the cross represent the fiducial cosmology, the BEAMS best fit and the  spectroscopic best fit respectively\cite{hlozek}. }
\label{fig:beams}
\end{figure}

Consider the following thought experiment: there is a room full of people and we desire to know the average height of women in the room. However, the only information we have is the height of each person and the length of their hair. Now, it is easy to imagine we could write down the probability that a person is a woman based on their hair length. How, using those probabilities can we estimate the average height of women? We could try applying a probability cut, for example assuming that anyone with a probability higher than 0.9 is a woman. This approach has some obvious problems: not only would we cut out much of the data (women with shorter hair) but also men with longer hair will be included and will bias the estimate of the average height. 
\\
This thought experiment is directly analogous to supernovae cosmology: we want to estimate cosmological parameters, such as $\Omega_m$ and $H_0$, from supernova data. We can obtain probabilities from the light curves of the  supernovae, but can only be sure of the type with spectroscopic follow-up.  Only the Ia's give cosmological information, like the women in the sample, but we cannot ignore possible contamination from the nIa's, as we could not ignore the long-haired men in the example.  Bayesian Estimation Applied to Multiple Species, or BEAMS\cite{kunz}, utilises all available data, correctly handles the contamination from the tall men and produces unbiased results. 
BEAMS includes the uncertainty in the types by marginalising over all possible combinations of types in the posterior:
\begin{equation}
P(\theta|D) = \displaystyle\sum\limits_{\tau} P(\theta,\tau|D),
\end{equation}
where $\theta$ is the set of cosmological parameters, $D$ is the data and $\tau$ is a length $N$ vector of types (where $N$ is the number of datapoints). The problem with this analytic marginalisation is that it is (assuming two types, Ia and nIa), an order $2^N$ calculation, which is computationally unfeasible. Fortunately, if the data are uncorrelated, Ref.~\refcite{kunz} showed that it is possible to simplify the calculation to order $2N$. The resulting calculation is essentially a kind of mixture model (see Ref.~\refcite{hogg} for an excellent video introduction).
\\
 Figure \ref{fig:beams} compares BEAMS with a probability cut and with the spectroscopic data only, when using both a controlled, simulated dataset and the SDSS-II dataset \cite{sdss1,sdss2}. The SDSS-II dataset is small, so no bias can be seen when a probability cut is used. However, as future surveys such as the LSST increase the number of supernovae dramatically, we expect a bias to become visible, similar to the constraints shown in panel (a).

\begin{figure}[th]
 \centering
\parbox{2.4in}{\epsfig{figure=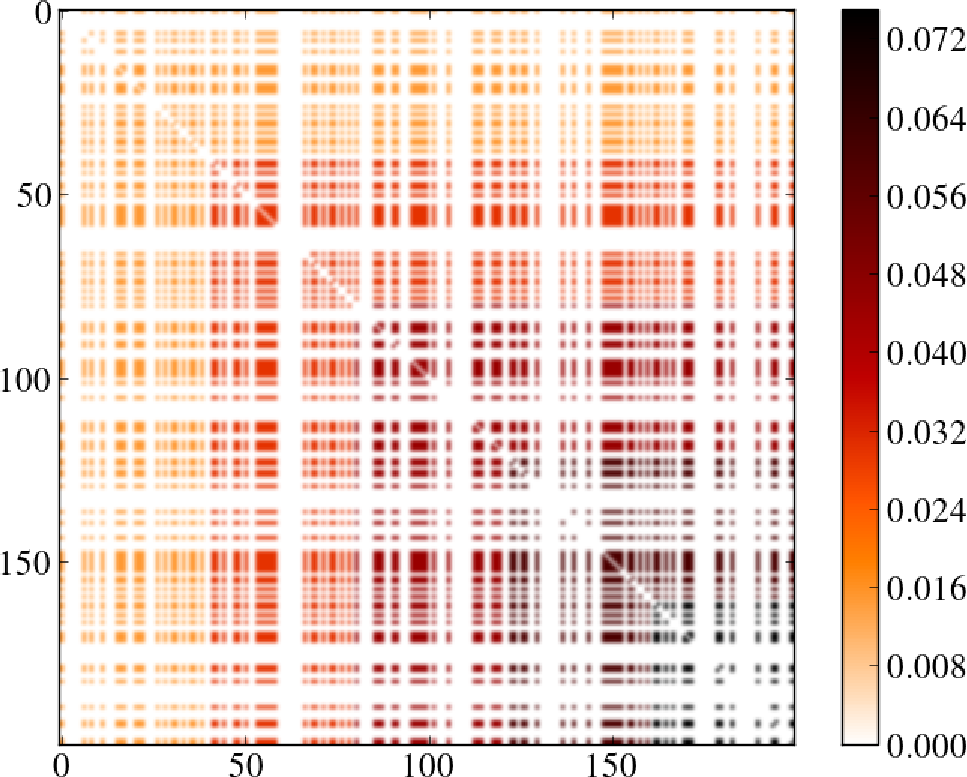,width=2.3in}
\figsubcap{a}}
 \hspace*{4pt}
 \parbox{2.4in}{\epsfig{figure=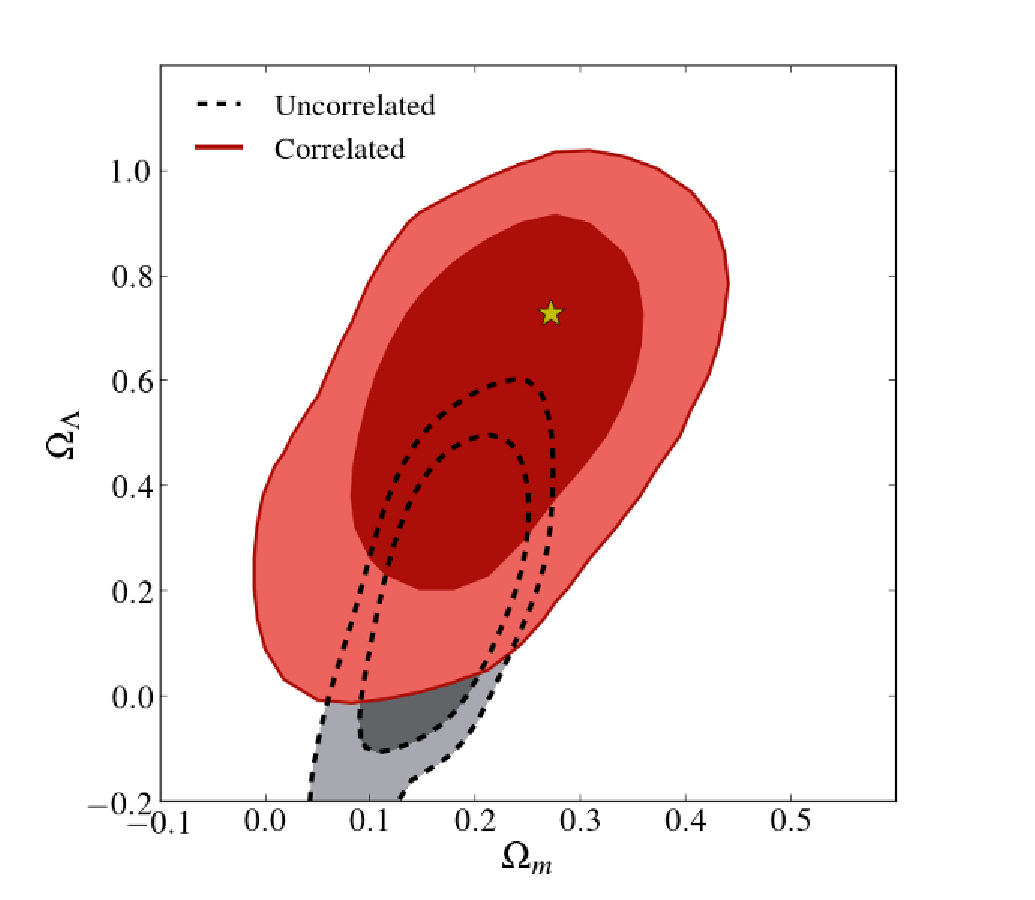,width=2.3in}
 \figsubcap{b}}
\caption{(a) An example covariance matrix\cite{kim}, where only the Ia's are correlated (and the diagonal has been removed for clarity). (b) $\Omega_m-\Omega_\Lambda$ contours produced when analysing the mock using the uncorrelated form of BEAMS (grey, dotted) and the correlated form with numerical marginalisation (red, filled).}
\label{fig:cor}
\end{figure}

Photometric supernovae data can have many sources of correlated systematic uncertainties, such as filter errors, peculiar velocities and redshift-stretch correlations\cite{knights}. If the uncorrelated form of BEAMS is used to analyse this data, biases can result. We developed a method \cite{knights} of using BEAMS for correlated data by marginalising over the types numerically instead of analytically. Figure \ref{fig:cor} shows an example covariance matrix used (a) with the corresponding $\Omega_m-\Omega_\Lambda$ contours (b), for a mock dataset of 200 strongly correlated supernovae. It is clear that without accounting for the correlated systematic uncertainties, it is possible to obtain biased contours which severely underestimate the errors on the parameters.
\vspace{-12pt} 
\vspace{-22pt}
\section*{Acknowledgments}
Michelle Knights acknowledges financial support provided by the University of Cape Town and the NRF/SKA, and is also grateful to Prof. Ariel Goobar and the  Department of Physics at Stockholm University for hosting her during MG13.

\vspace{-7pt}
\bibliographystyle{ws-procs975x65}
\bibliography{beams_refs}

\end{document}